\documentclass{mn2e}

\input{psfig}
\newcommand{\beq}{\begin{equation}}
\newcommand{\eeq}{\end{equation}}

\def\gtsima{$\; \buildrel > \over \sim \;$}
\def\ltsima{$\; \buildrel < \over \sim \;$}
\def\prosima{$\; \buildrel \propto \over \sim \;$}
\def\gsim{\lower.7ex\hbox{\gtsima}}
\def\lsim{\lower.7ex\hbox{\ltsima}}
\def\simgt{\lower.7ex\hbox{\gtsima}}
\def\simlt{\lower.7ex\hbox{\ltsima}}
\def\simpr{\lower.7ex\hbox{\prosima}}
\def\la{\lsim}
\def\ga{\gsim}

\newcommand{\apjs}{ApJS}

\newcommand{\araa}{ARA\&A}

\newdimen\hssize
\newdimen\hdsize
\title[Excess Far-IR Emission of AGN]
      {The Excess Far-Infrared Emission of AGN in the Local Universe}               

\author[A.~Pasquali et al.]{A.~Pasquali,$^1$\thanks{Email:
pasquali@phys.ethz.ch}
G.~Kauffmann,$^2$  T.M. Heckman,$^3$\\  
$^1$Institute of Astronomy, ETH Hoenggerberg, HPF, 8093 Zurich, Switzerland\\
$^2$Max-Planck-Institut f\"ur Astrophysik, Karl Schwarzschild Str. 1, 85748
Garching, Germany\\
$^3$Center for Astrophysical Sciences, Department of Physics and Astronomy, 
Johns Hopkins University, Baltimore, MD, 21218 USA\\}

\begin{document}

\date{}

\pagerange{\pageref{firstpage}--\pageref{lastpage}} \pubyear{2002}

\maketitle

\label{firstpage}

\begin{abstract}
We have cross-correlated  the Sloan Digital Sky Survey (SDSS) 
second data release spectroscopic
galaxy sample with the IRAS faint-source catalogue (FSC).
Optical emission line
ratios are used to classify the galaxies with reliable IRAS
60 and 100 $\mu$m detections  into AGN and normal star-forming galaxies.  
We then create  subsamples of normal galaxies and AGN that are very closely    
matched in terms of key physical properties such as stellar mass, redshift,
size, concentration and mean stellar age (as measured by absorption
line indicators in the SDSS spectra).
We then quantify whether
there are systematic differences between the  IR luminosities of the
galaxies and the AGN in the matched subsamples. We find that the AGN exhibit
a significant excess in  far-IR 
emission relative to the star-forming galaxies in our sample. 
The excesses at 60 $\mu$m and 100 $\mu$m are $0.21 \pm 0.03$ dex
and $0.12 \pm 0.035$ dex in $\log L_{60}/M_{\star}$ and 
$\log L_{100}/M_{\star}$, respectively.
We then discuss whether the far-IR excess is produced by 
radiation from the active nucleus that is absorbed by dust or alternatively,
by an extra population of young stars that is not detectable at optical
wavelengths.
\end{abstract}


\section{Introduction}

In the mid  1980's, spectroscopic follow up of 
galaxies observed by the IRAS satellite revealed
that infrared-luminous galaxies consist of a mixture of star-forming       
galaxies and galaxies with an active nucleus (AGN).
De Grijp et al. (1985) found that fraction of AGN could be 
maximized by selecting IRAS sources with relatively warm 25 to 
60 micron colours and they then used this criterion to construct a catalog
of 80 Seyfert 1 and 141 Seyfert 2 galaxies (De Grijp et al. 1992).

A considerable amount of effort has been devoted to understanding
the physical mechanisms responsible for the infrared emission
in AGN. The current paradigm asserts that type 1 and type 2 
Seyfert galaxies are drawn from the same parent population.
In type 1 Seyfert galaxies the subparsec-scale continuum
source is viewed directly, but in type 2 Seyferts this
source is blocked from view by a structure (commonly referred to
as the ``torus'') with a size of between 1 and 10 parsec, which is
optically thick to radiation from X-ray to near-IR wavelengths.
The size of the obscuring region, together with energy conservation
arguments, then imply that most of the radiation absorbed by this
structure will be re-radiated in the mid-IR as thermal emission
from dust (Pier \& Krolik 1992; Granato \& Danese 1994).

In recent years it has become clear that star formation and AGN activity
frequently occur together in galaxies (e.g. Cid Fernandes et al. 2001;
Kauffmann et al. 2003c; Heckman et al. 2004). As a result, it is
natural to suppose that a substantial fraction of IR emission from
AGN may be produced by star formation. It is commonly believed
that star formation is likely to dominate the emission
at long wavelengths ($> 60 \mu$m)  
and that  emission from the
torus prevails in the mid-IR.
Rowan-Robinson \& Crawford (1989) modelled the far-IR spectra of
IRAS galaxies using three components: a ``disc component'' due to the
interstellar dust illuminated by the galaxy's starlight; a ``starburst
component'' arising from hot stars in optically thick dust clouds,
and a ``Seyfert component'' due to a power-law continuum source within
the torus. In most  cases ($\sim$ 61$\%$),
the IRAS spectra are well fitted by a combination of disc and 
starburst components. Only $\sim$ 24$\%$ of the galaxies in the samples
required the Seyfert component. 
Nevertheless, as discussed by
Silva, Granato \& Maiolino (2004), the intrinsic far-IR properties of AGN              
remain subject to strong uncertainties (only large-aperture
data from ISO or IRAS are available at $\lambda > 20 \mu$m)
and there is also substantial freedom in the dusty torus
models at these wavelengths.                               

The relative importance of dust heating by star formation and by
AGN has also been a topic of considerable controversy  in the study
of ultraluminous IRAS galaxies (ULIRGs), which are the most powerful 
galaxies detected by IRAS with IR luminosities in
excess of $10^{12} L_{\odot}$. A significant fraction of ULIRGs  
exhibit nuclear optical emission line spectra characteristic of
Seyfert galaxies (Sanders et al. 1988; Armus et al. 1990),
but the infrared, millimeter and radio characteristics of these
systems are very similar to those of ordinary starbursts
(Rieke et al. 1985; Rowan-Robinson \& Crawford 1989; Condon et al. 1991). 
Genzel et al. (1998) used mid-infrared spectra from the ISO satellite
to demonstrate that 70-80\% of the ULIRGS in their sample  
were predominantly powered by star formation and 20-30\% by a central 
AGN. These conclusions were based on an analysis of the
ratio of high- to  low-excitation mid-IR emission lines
as well as the strength of the 7.7 $\mu$m PAH feature in these systems.

The same question is relevant when attempting to convert from the
luminosity function of sub-millimeter sources observed at
high redshifts to estimates of the total star formation rate
density (e.g. Blain et al. 1999; Rowan-Robinson 2001). The contribution
from AGN has been investigated by looking for overlap 
between sub-mm galaxies detected by 
SCUBA and hard X-ray sources found  by Chandra. Only around 10\%
of the sub-mm sources are  found to have an X-ray counterpart
(Severgnini et al. 2000; Barger et al. 2001). More recently Silva et al. (2004)
have used a combination of X-ray data on the evolution of the AGN
luminosity function and spectral energy distributions drawn from
their models to show that around 95\% of the total IR background
is likely produced by star formation.

In this paper, we study the far-infrared properties of AGN
in the local Universe.
As a result of recent  large redshift surveys 
such as the Sloan Digital Sky Survey (SDSS), it is possible to compile
unprecedentedly large samples of galaxies with both high quality optical spectra   
and IRAS fluxes. It then becomes possible to adopt 
a purely statistical approach to understanding the origin of the IR emission
in AGN. In this paper, we analyze a sample of over a thousand  galaxies 
drawn from the SDSS Data Release 2 (DR2) with
IRAS detections. 
Optical emission line
ratios are used to classify the galaxies into AGN and ``normal'' galaxies.  
We then create  subsamples of normal galaxies and AGN that have been carefully
matched in terms of key physical properties such as stellar mass, redshift,
galaxy structural parameters and mean stellar age, and we quantify whether
there are systematic differences between the mean IR luminosities of the
galaxies in the matched subsamples.

\section{The  galaxy samples}

\subsection{The SDSS Spectroscopic Sample}

The Sloan Digital Sky Survey (York et al. 2000; Stoughton et al. 2002, and
references therein) is an optical imaging (u,g,r,i,z
bands) and spectroscopic survey of about a quarter of the extragalactic
sky, being carried out at the Apache Point Observatory.  The spectroscopic
sample considered in this paper is a sample of about 212,000 objects with
magnitudes $14.5 < r < 17.77$, spectroscopically confirmed to be galaxies.
This sample of galaxies is
described by Brinchmann et al. (2004) . The galaxies have a
median redshift of $z \sim 0.1$.

The SDSS spectra cover an observed wavelength range of 3800 to 9200\AA, at
an instrumental velocity resolution of about 65km\,s$^{-1}$. The spectra
are obtained through fibres of about 3-arcsecond diameter, which corresponds
to 5.7 kpc at a redshift of 0.1; at this redshift the spectra therefore
represent a large proportion (up to 50\%) of the total galaxy light,
whilst for the very lowest redshift objects they are more dominated by the
nuclear emission.

As described by Brinchmann et al. (2004), many properties of
these galaxies have been parameterised, with the derived catalogues of
parameters being publically available on the web. Derived parameters
include: fundamental galaxy parameters such as total stellar masses,
sizes, surface mass densities, concentration indices, mass-to-light
ratios, 4000\AA\ break strengths, dust attenuation measurements, and
H$\delta$ absorption measurements (Kauffmann et al. 2003a) ; accurate emission
line fluxes, after subtraction of the modelled stellar continuum to
account for underlying stellar absorption features (Kauffmann et al. 2003a;
Tremonti et al. 2004); galaxy metallicities (Tremonti et al. 2004);
star formation rates (Brinchmann et al. 2004);
parameters measuring optical AGN activity, such as emission line ratios,
and galaxy velocity dispersions (hence black hole mass estimates; Kauffmann
et al. 2003c, Heckman et al. 2004). The sample analysed in this paper
includes star-forming galaxies and those AGN in which
non-stellar continuum light from the nucleus has a negligible effect on the
physical parameters derived for the host galaxy (see Kauffmann et al. 2003c
for more detailed discussion). In the rest of the paper we use the term
``AGN'' to refer to these objects.
\footnote{Note that this sample explicitly excludes those objects
classified by the SDSS spectroscopic pipeline as
QSOs. In such cases the SDSS spectrum is dominated by light from the AGN.}

\subsection{Cross-identification with the IRAS 
catalogues}

We cross-identified the SDSS galaxies
with the IRAS Faint-Sources
Catalogue (FSC) using the web search engine GATOR (available at 
http://irsa.ipac.caltech.edu/applications/Gator). 
We initially assumed a search radius of 1 arcmin in order to
account for the IRAS positional uncertainties.
Initially, no constraint  on the IRAS flux density or the quality
of the IRAS flux detections was applied and the query returned
a total of  5765 matches. Because of the large search radius,
a substantial number of IRAS sources (1205 or 21\%) were matched to more
than one  SDSS galaxy. These sources were
eliminated from our catalogue in order to maximize the reliability
of our sample. 
Our catalogue is thus incomplete and somewhat biased
against galaxies in high density regions, but this is not
particularly important  for the applications discussed in this paper.

We restricted our analysis to IRAS sources with 
reliable flux densities [with flux quality from
moderate (2) to good (3), which means that they
have been detected more than once in the repeat scans].
As discussed in detail
in Kauffmann et al. (2003c), we use the [NII]/H$\alpha$
versus [OIII]/H$\beta$  emission line ratio diagnostic diagram
(Baldwin, Phillips \& Terlevich 1981) to classify our galaxies
into AGN and normal galaxies. In order for a galaxy to be placed on the BPT
diagram, the four emission lines [NII],[OIII],H$\alpha$ and H$\beta$ 
must  be detected with S/N$>3$. Galaxies in which these four lines
are not detected with sufficient signal-to-noise
are classified as ``normal''.  The normal galaxy sample 
may thus contain some  AGN with emission lines that are too weak to classify. 
We are not concerned with this here,
because the aim of this paper is to characterize the infrared emission from
more powerful Seyfert galaxies.
Figure 1 shows the distribution in the BPT diagram of the 
emission-line galaxies in our sample with
 60 and 100 $\mu$m  IRAS detections.
Red crosses indicate galaxies that are classified
as AGN. As can be seen, 
most of the AGN in our sample
lie in the region of the diagram occupied by Seyfert
galaxies or by  ``composite'' systems in
which there is both an active nucleus and ongoing star formation
(see Kauffmann et al. 2003c for a more detailed discussion
of how the general population of AGN in the SDSS populate the BPT diagram). There
are almost no IRAS-selected AGN in the region of the diagram
occupied by LINERs.

\begin{figure*}
\centerline{\psfig{figure=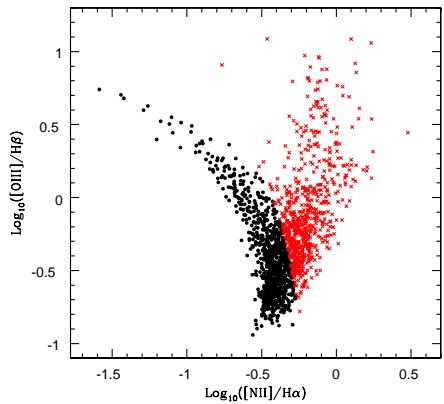,width=0.7\hssize}}
\caption{The distribution of IRAS-selected  emission-line galaxies in 
the [NII]/H$\alpha$ versus [OIII]/H$\beta$ BPT diagram. Red crosses 
indicate objects that are classified as AGN of Type 2.}
\end{figure*}

The infrared luminosities of the galaxies
in our sample were computed using the formulae given in
Helou et al. (1988) and Sanders \& Mirabel (1996):
\vskip 0.2truecm
\par\noindent
FIR = 1.26 $\times$ [F(60$\mu$m) $+$ F(100$\mu$m)] W~m$^{-2}$
\vskip 0.15truecm
\par\noindent
F(60$\mu$m) = 2.58 $\times$ 10$^{-14}f_{\nu}$(60$\mu$m) W~m$^{-2}$
\vskip 0.15truecm
\par\noindent
F(100$\mu$m) = 1.00 $\times$ 10$^{-14}f_{\nu}$(100$\mu$m) W~m$^{-2}$
\vskip 0.15truecm
\par\noindent
L$_{FIR}$ = 4$\pi$D$^2_L$FIR 
\vskip 0.15truecm
\par\noindent
L$_{100}$ = 4$\pi$D$^2_L$F(100$\mu$m)
\par\noindent
L$_{60}$ = 4$\pi$D$^2_L$F(60$\mu$m)
\vskip 0.2truecm
\par\noindent
where  $f_{\nu}$(60$\mu$m) and
$f_{\nu}$(100$\mu$m) are the IRAS flux densities and we have adopted
a cosmolgy with  H$_o$ = 70 
km~s$^{-1}$~Mpc$^{-1}$, $\Omega=0.3$ and $\Lambda=0.7$ 

\begin{figure*}
\centerline{\psfig{figure=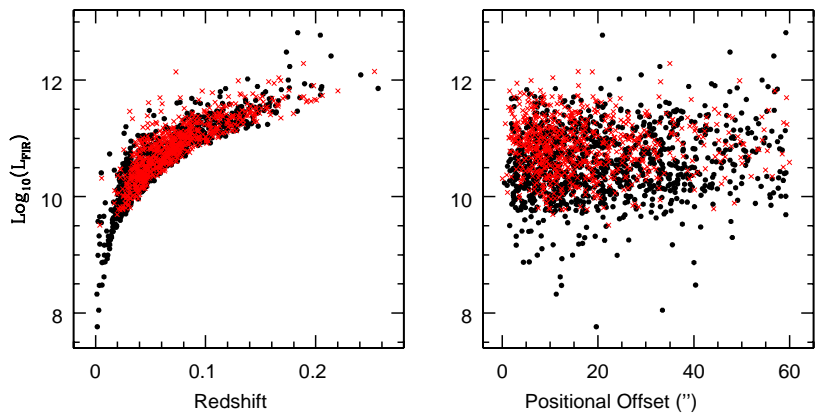,width=1.3\hssize}}
\caption{Left: Log(L$_{FIR}$) is plotted as a function of redshift for
AGN of Type 2 (red crosses) and normal star-forming galaxies 
(black points).
Right: Log(L$_{FIR}$) is plotted as a function of the separation 
(in arcseconds) between the IRAS source and the matching SDSS galaxy. 
L$_{FIR}$ is in solar luminosity units.}
\end{figure*}

The left panel of Figure 2 shows the FIR luminosities (in solar units) 
plotted as a function of redshift for all the galaxies with reliable
flux measurements at both 60 and 100 $\mu$m.
The  IRAS flux limit means that the
IR luminosities of the objects in our sample increase from 
L$_{FIR}$ $\simeq 10^{10} L_{\odot}$  at $z$ = 0.05 to
$\sim 10^{11} L_{\odot}$ at $z=0.1-0.2$. This is true for both normal 
galaxies and for  AGN. In the right panel of Fig. 2, we plot
the derived FIR luminosity as a function of the distance in arcseconds between
the position of the IRAS source and the position of the matching SDSS
galaxy.

\subsubsection {Further checks on the reliability of the sample}

In our sample, the signal-to-noise of the
detected sources
ranges from 3.5  up to 50. 
Follow-up of IRAS FSC sources have shown that
at $S/N < 8$, a significant fraction (more than 20\%) of the
sources turn out to be false detections. Because of the large
matching radius that is  employed, one might worry that
some of these false sources may contaminate our sample.                  
Another source of potential contamination is cirrus emission from
interstellar dust in our own Galaxy. In the IRAS catalogue, this
is parametrized by the ``cirrus flag'', which indicates the number of
100 $\mu$m sources detected within a 30 arcmin radius
of the FSC source.

Figure 3 shows the distribution of positional offsets (in arcseconds)
between the IRAS FSC source and the matched SDSS  galaxy
for different cuts in S/N and in cirrus flag. As 
the S/N of the sources increases there is a slight shift towards
smaller offsets, indicating that some false identifications
have been eliminated. The shifts are quite small, however, 
indicating that the majority of low S/N objects are in fact real.
Different cuts in the cirrus flag have
no effect on the distribution of offsets.

We have generated catalogues of randomly distributed 
sources in the area of sky
covered by DR2   and we use these to evaluate the likelihood that 
a false detection will be matched with an SDSS galaxy  
from the main spectroscopic sample. Results are shown in 
Figure 4 as a function of the matching radius. 
Since, within the footprint covered by DR2, the FSC contains 4560
sources associated with galaxies, Figure 4 indicates that about 288
false matches occur for a matching radius
of 60 arcseconds. At a 30 arcsecond matching radius,
the number of randomly matched galaxies drops to about 46, 
clearly a more acceptable level of contamination.
From Figure 3 we see that a cut at 
at a positional offset of 30 arcseconds eliminates only around a quarter
of the galaxies in our sample. We conclude that
a cut in positional offset is a more efficient way to eliminate 
contaminating sources than a cut in signal-to-noise.

\begin{figure*}
\centerline{\psfig{figure=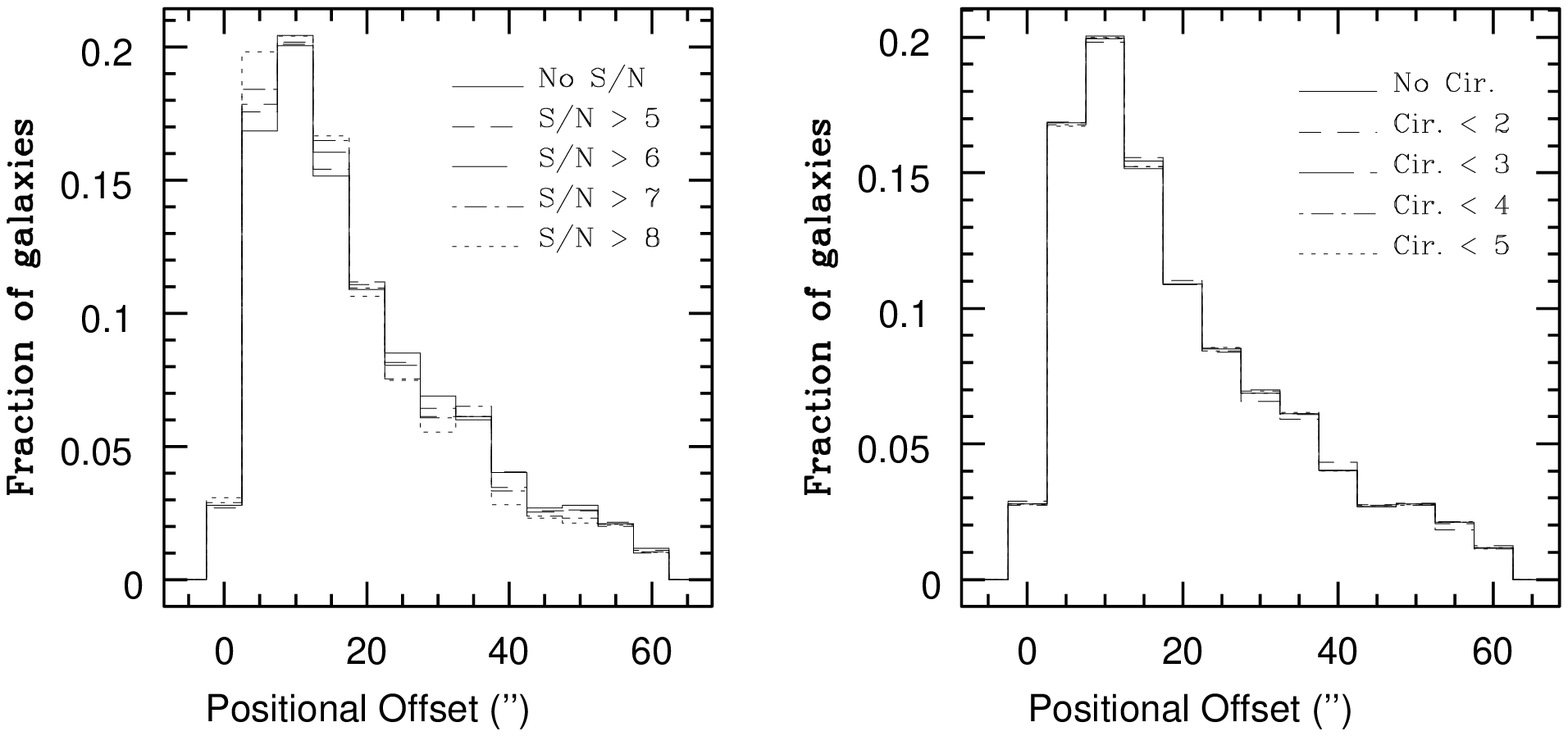,width=1.2\hssize}}
\caption{ The distribution of positional offset between the FSC
source and the matched SDSS galaxy for different cuts in S/N (left)
and cirrus flag (right).}
\end{figure*}

\begin{figure*}
\centerline{\psfig{figure=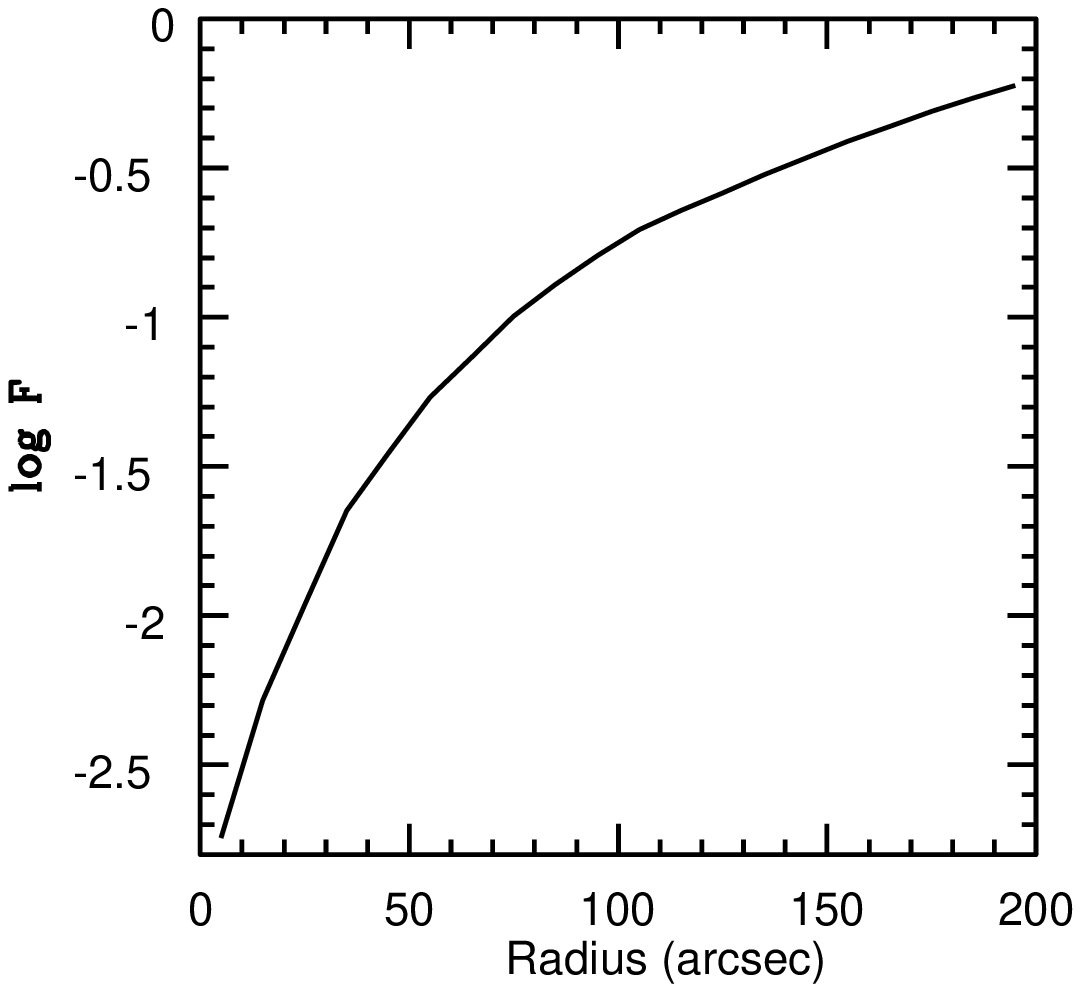,width=0.7\hssize}}
\caption{The logarithm of the fraction of radomly placed sources that     
are matched to an SDSS galaxy as a function of the matching radius.}
\end{figure*}

Another way to evaluate the reliability of the sample is to
study  the scatter in the relations between 
properties of galaxies derived from the SDSS spectra and those
derived from the IRAS fluxes.
In Figure 5 we study  the effect of different cuts in
positional offset and in redshift on the relation between the
4000 \AA\ break strength D$_n$(4000) and the ``normalized'' FIR luminosity
$L_{\rm FIR}/M_*$. Both quantities are a measure of the age
of the stellar population in the galaxy and will be used extensively
in the analysis that follows. 

In the top panels of Fig. 5, we hold the positional offset cut
constant at R $<$ 30 arcseconds and we show what happens if different
cuts in redshift are imposed on the sample.
At high redshifts , the sample contains only 
extremely IR-luminous objects. Fig. 1 shows
there are a handful of galaxies with IR luminosities in excess of
$10^{12} L_{\odot}$ at $z>0.15$. These are the ULIRGs, which have         
been traditionally considered as a separate class of galaxy in their own right.
Since there are very  few of these objects in our sample,
we exclude them by imposing a  cut at $z=0.15$.

At low redshifts the SDSS spectra, which are obtained using
3 arcsecond diameter fibres, sample only the  inner regions
of the galaxies and aperture effects may  influence
our analysis. Kewley, Jansen \& Geller (2005) have recently
carried out a detailed study of aperture effects on the
spectra of galaxies using a sample of 101 nearby galaxies
with both global and nuclear spectra. They conclude that so
long as the fibre captures more than $\sim$ 20\%  of the total
light, the differences between physical parameters derived
using nuclear spectra and those derived using global spectra
are modest. For SDSS, Kewley et al.  recommend that redshift
cut $z > 0.04$ be imposed. We choose a somewhat more  
conservative cut  ($z>0.06$). The top panels
in Fig. 5 show that this cut eliminates a population of ``normal''
galaxies with  high values of $L_{\rm FIR}/M_*$. These
are primarily low mass galaxies ($< 10^{10} M_{\odot}$) 
that are currently experiencing  a strong
burst of star formation. There are very few AGN that have 
stellar masses lower than $10^{10} M_{\odot}$ (Kauffmann et al. 2003c), 
so eliminating these objects from our sample
will not affect the analysis presented in this paper.

In the bottom panels, we fix the redshift interval
at $0.06 < z < 0.15$ and we investigate what happens
if the cut in positional offset is allowed to vary. 
As can be seen, if matches with
positional offsets as large as 60 arcseconds are accepted, there is a ``tail''
of outlying galaxies with large 4000 \AA\ break strengths (indicative
of an old stellar population), but with high normalized FIR luminosity.
These outliers largely disappear when the offset is restricted to
be less than 30 arcseconds.

\begin{figure*}
\centerline{\psfig{figure=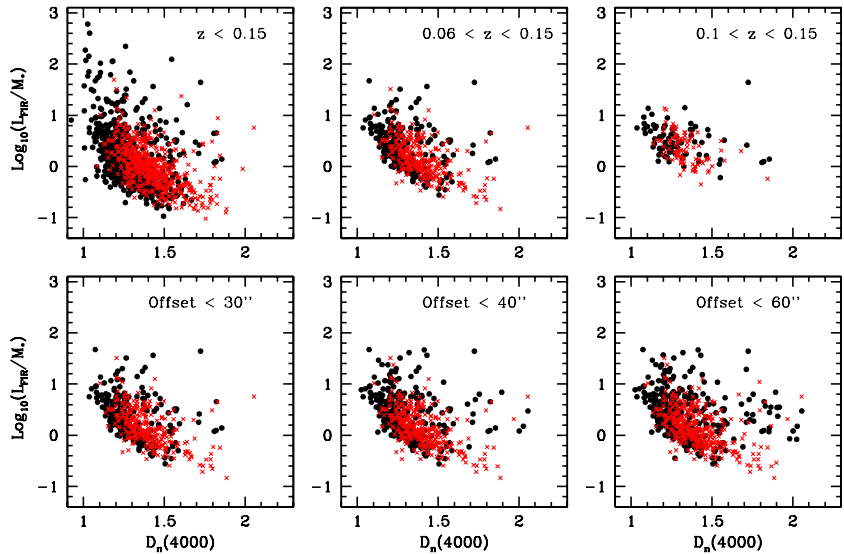,width=1.35\hssize}}
\caption{Top: The relation between normalized IR luminosity and 4000 \AA\
break strength for different cuts in redshift. The positional
offset is fixed to be $ R<$ 30 arcseconds. Bottom: The relation
is shown for different cuts in positional offset while holding
the redshift cut fixed at $0.06 < z < 0.15$.
Red crosses indicate AGN of Type 2
and black dots normal star-forming galaxies.}
\end{figure*}

To summarize,  the final cuts we impose on our sample are the following:
\vskip 0.3truecm
\par\noindent
\par\noindent
(i) Only IRAS sources with flux quality flag $\ge$ 2  are retained.
\par\noindent
(ii) The positional offset between the IRAS source and the SDSS
galaxy must be less than 30".
\par\noindent
(iii) All sources that have more than one SDSS match within 
a matching radius of 30" are eliminated.
\par\noindent
(iv) All galaxies with $z<0.06$ and $z>0.15$ are also excluded.
\vskip 0.3truecm
\par\noindent
We created two catalogues: one with flux measurements
at 60 $\mu$m only  (1090 galaxies of which 553 are AGN) and a second
with flux measurements at both 60 and 100 $\mu$m 
(526 objects of which 284 are AGN). All sources in the 60 $\mu$m
catalogue have an IRAS flux quality of 3, while all galaxies in the 100
$\mu$m are characterised by a quality of 2.

\section{Properties of the IRAS-selected AGN }

In Figure 6 we show the distributions of some of the                           
basic properties of the AGN and the normal star-forming galaxies in our
sample, such as their redshifts, IR luminosities,
stellar masses, concentrations, stellar surface densities and
4000 \AA\ break strengths. Unless specified otherwise,
we show results for the catalogue with reliable flux
measurements at both 60 and 100 $\mu$m.
The histograms representing the normal  galaxies and the 
AGN have been  shaded in black and in red,
respectively. 

\begin{figure*}
\centerline{\psfig{figure=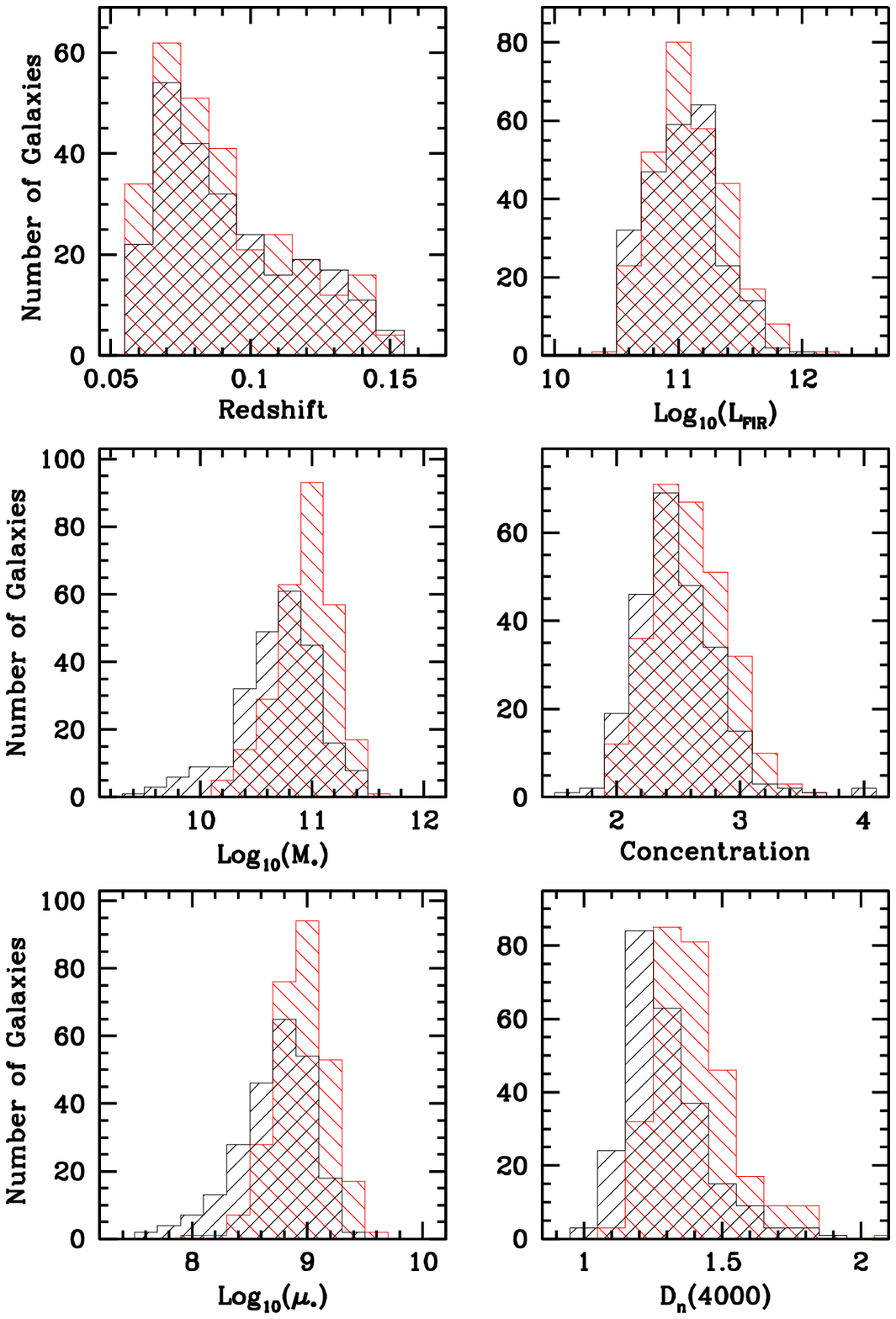,width=1.1\hssize}}
\caption{The distribution of IRAS-selected galaxies as a 
function of redshift, $\log L_{ FIR}$, stellar mass $M_*$,
concentration index $C$, stellar surface mass density $\mu_*$
and 4000 \AA\ break strength D$_n$(4000).
The histograms of normal star-forming galaxies and AGN of
Type 2 are shaded in 
black and in red, respectively.}
\end{figure*}

The main result seen in Figure 6 is that the IRAS-selected AGN have larger
stellar masses than the ```normal'' galaxies in the sample.
As discussed by Kauffmann et al. (2003c), the AGN fraction among
normal galaxies falls off  very steeply
at stellar masses less than  a few $ \times 10^{10} M_{\odot}$
and this is apparent in the third panel of Figure 6.
As discussed in Kauffmann et al. (2003b), more massive galaxies
have higher concentrations and surface densities and their
stellar populations are also older. It is therefore
not surprising that Figure 6 shows that the AGN  are biased
to higher values of $C$, $\mu_*$ and D$_n$(4000) when compared to
the star-forming galaxies.

In Figure 7, we plot                                     
the ``normalized''
IR-luminosity  $L_{FIR}/M_*$
as a function of stellar mass, concentration, stellar surface mass density
and 4000 \AA\ break strength. Once again solid black points indicate 
normal galaxies and red crosses AGN.  It should be noted that the strong  correlation
between $L_{FIR}/M_*$ and  stellar mass and surface density is
a selection effect caused by our redshift cut and by the 
IRAS flux detection limit.
However, it possible to conclude from this plot that the
{\em largest} normalized IR-luminosities are obtained for galaxies 
with the lowest masses, concentrations and surface densities. 
There are  few AGN in this region of parameter space. 
At a {\em fixed} value of $M_*$, $C$
or $\mu_*$, however, the differences between AGN and non-AGN
are much more subtle. There does not appear to be a significant difference
between the normalized IR-luminosities of AGN and non-AGN at
fixed stellar mass or concentration, but AGN do appear to be offset to
slightly higher values of $L_{\rm FIR}/M_*$ at a fixed value of D$_n$(4000).

\begin{figure*}
\centerline{\psfig{figure=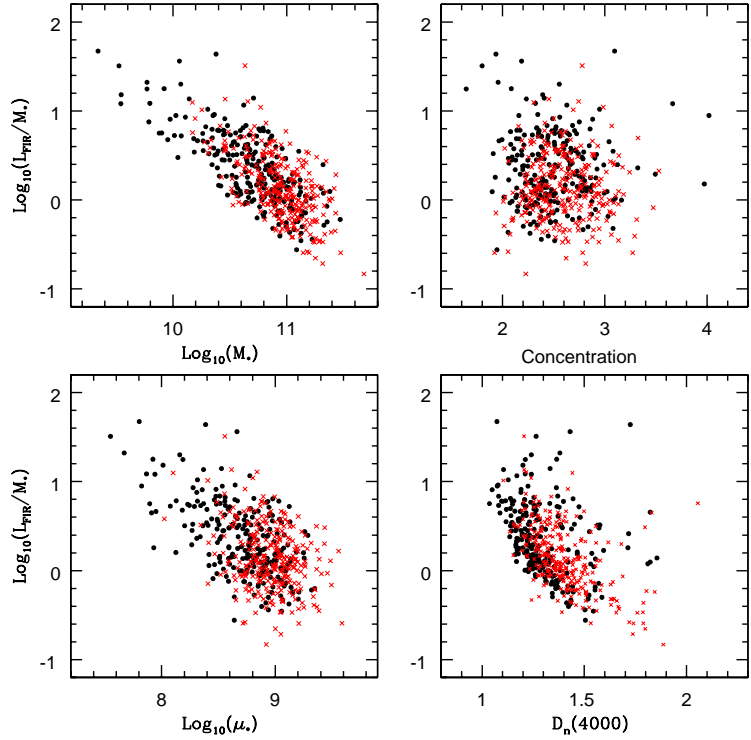,width=1.2\hssize}}
\caption{The normalized IR luminosity $L_{\rm FIR}/M_*$ is plotted
as a function of stellar mass, concentration, stellar surface
mass density and 4000 \AA\ break strength for both normal star-forming
galaxies (black points) and for AGN of Type 2 (red crosses). }
\end{figure*}

We now divide our samples of  AGN and normal star-forming galaxies (non-AGN)
into bins of stellar mass with a 
width of $\Delta$Log$_{10}$(M$_*$) = 0.2. We only consider bins that contain
at least 10 AGN and 10 normal star-forming galaxies. 
We compute the mean Log$_{10}$(L$_{FIR}$/M$_*$) of the AGN and the
normal star-forming galaxies
in each bin and we define the ``AGN excess''  
$\Delta$Log$_{10}$(L$_{FIR}$/M$_*$) as 
\par\noindent
$<$Log$_{10}$(L$_{FIR}$/M$_*$)$>_{AGN}$ - 
$<$Log$_{10}$(L$_{FIR}$/M$_*$)$>_{non-AGN}$
\par\noindent
To analyze whether our results depend on wavelength,
we  have also calculated $\Delta$Log$_{10}$(L$_{100}$/M$_*$)
and $\Delta$Log$_{10}$(L$_{60}$/M$_*$), which use only the 100 or 60 $\mu$m
fluxes rather than the combination of the two quantities.
Our results are shown in Figure 8.
The error bars have been computed using  standard bootstrap resampling
techniques.
As can be seen, at a fixed stellar mass 
 there is very little difference in normalized
IR-luminosity between AGN and normal star-forming galaxies at any wavelength.

\begin{figure*}
\centerline{\psfig{figure=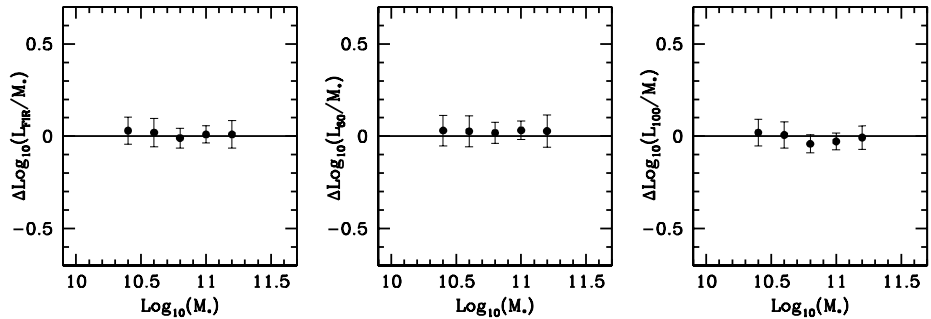,width=1.5\hssize}}
\caption{The difference between the average normalized IR luminosity 
of AGN of Type 2 and the average normalized IR luminosity of normal
star-forming galaxies
is calculated in bins of stellar mass $M_*$. Results are shown 
for $L_{\rm FIR}$ and well as for $L_{60}$ and $L_{100}$.}
\end{figure*}

In the next step of the analysis, we divide our sample into bi-dimensional
bins in stellar mass {\em and} D$_n$(4000).
Once again, we only include bins
with more than 10 AGN and normal star-forming galaxies in our analysis. 
The adopted bin sizes are 0.5 dex
in $\log M_*$ and 0.1 in D$_n$(4000). The left hand panels of
Figure 9 show histograms
of the distribution of
$\Delta$Log$_{10}$(L$_{FIR}$/M$_*$),  
$\Delta$Log$_{10}$(L$_{60}$/M$_*$),  
and $\Delta$Log$_{10}$(L$_{100}$/M$_*$) for the bins with sufficient
galaxies and AGN  to perform the comparison. In the right hand panels, the
same quantities are plotted as a function of the value of D$_n$(4000)
at the center of each  bin. We conclude that if galaxies and AGN are
matched in both stellar mass and 4000 \AA\ break strength, AGN
tend to be brighter than star-forming galaxies by an amount which
is larger at 60 $\mu$m than at 100 $\mu$m. However, this infrared
excess is significant at a 2$\sigma$ level at 60 $\mu$m and only at
1$\sigma$ level at 100 $\mu$m and in L$_{FIR}$.
The reason the excess is not seen in  
Figure 8 is because AGN are found in  galaxies with 
larger 4000 \AA\ break strengths than star-forming galaxies of the same mass.

\begin{figure*}
\centerline{\psfig{figure=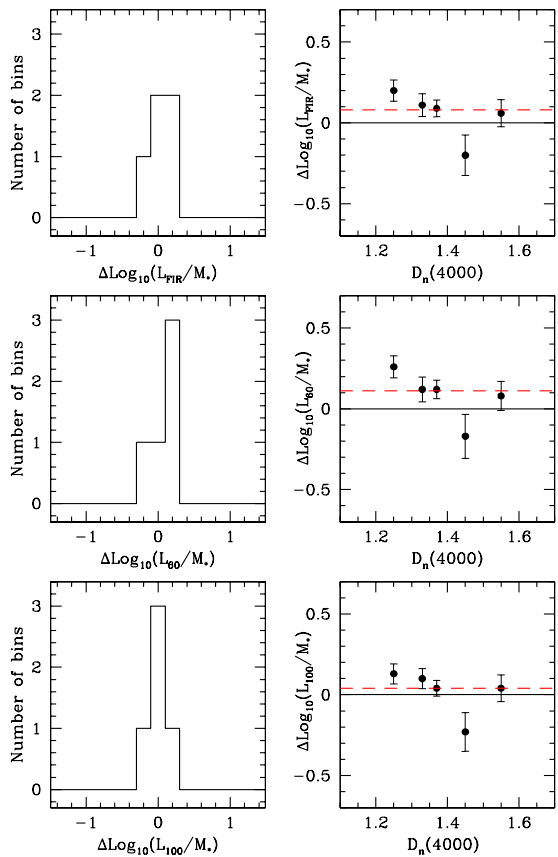,width=0.9\hssize}}
\caption{The difference between the average normalized IR luminosity 
of AGN of Type 2 and the average normalized IR luminosity of normal 
star-forming galaxies
is calculated  in bins of stellar mass and D$_n$(4000) . The left
panels show histograms of the offset values and in the right panels,
the offsets are plotted as a function of D$_n$(4000) at the center of
the bin. Results are shown for 
$L_{\rm FIR}$,  $L_{60}$ and $L_{100}$. The dashed red line
indicates the average value of offset. The errorbars correspond to
1$\sigma$.}
\end{figure*}

\section{Matched Pair Analysis}

The differences between Figures 8 and 9 teach us that 
in order to compare the far-IR 
properties of AGN and normal star-forming galaxies in an unbiased way, 
it is important to match the properties of
the host galaxies of these two kinds of  systems as 
closely as possible. The relatively small number
of galaxies in our sample means that it is 
not feasible to bin in more than two
dimensions.  Instead, we choose to create a sample of galaxy-AGN
{\em pairs} that are closely matched in stellar mass $M_*$, D$_n$(4000),
concentration index $C$, stellar surface mass density $\mu_*$ (hence physical
size) and redshift $z$. In order to maximize the number of pairs, we first   
analyze the  catalogue with  reliable flux measurements at 60 $\mu$m.
We accept pairs if $\Delta \log M_* < 0.25$, $\Delta$D$_n$(4000)$< 0.04$,
$\Delta C < 0.1$, $\Delta \log \mu_* < 0.25$ and $\Delta z <$ 0.03. This leaves us
with a sample of 254 unique galaxy-AGN pairs.

In Figure 10, we plot $\Delta \log L_{60}/M_*$ 
(the difference between the normalized
60 $\mu$m luminosity of the AGN and the matched galaxy ) for each pair as a
function of a number of  parameters describing the AGN. 
The red points on the plot
show the average value of $\Delta \log L_{60}/M_*$
calculated in bins of each parameter (the red point  is positioned
at the center of the bin). The close pair analysis indicates that
there is a 0.2 dex excess
in $\Delta \log L_{60}/M_*$ for the AGN
relative to the normal star-forming galaxies. The excess does not 
depend on redshift (showing that
is not caused by aperture effects) or on  structural properties 
such as galaxy concentration or  stellar surface mass density.
It does appear to be larger for lower mass  AGN  with smaller 4000 \AA\
breaks  and for more powerful AGN with larger  extinction-corrected
[OIII] luminosities.

In order to assess the error in the measured far-IR excess of AGN,
we generated 5 different galaxy/AGN pair samples by starting the
search for pairs from different points in the catalogue.
In addition, we  created 300 bootstrap resamplings of each of the
5 pair samples in order to assess whether our results were
sensitive to the presence of a few outliers in the distribution.
The derived 60 $\mu$m excess for each of the 5 samples
and the associated error estimated from the bootstrap resamplings are 
listed in Table 1. As can be seen, the variance among the 5 samples is 
consistent with the errors calculated using the bootstrap technique and is 
around 0.03 dex. We thus conclude that our measured 60 $\mu$m excess of 0.2 
dex is statistically significant. In the same way, we derived an error
of 0.035 dex on the excess measured at 100 $\mu$m and in L$_{FIR}$.
This implies that the excesses found at 60 and 100 $\mu$m differ at
about the 2$\sigma$ level.

\begin{table}
\caption{The results of the bootstrap technique applied to the
60 $\mu$m catalogue.}
\begin{tabular}{ccc}
\hline
Sample & Excess & 1$\sigma$ error\\
       & at 60 $\mu$m &\\
       \hline
1 & 0.208 & 0.0327 \\
2 & 0.174 & 0.0213 \\
3 & 0.205 & 0.0331 \\
4 & 0.180 & 0.0194 \\
5 & 0.204 & 0.0324 \\
\hline
\end{tabular}
\end{table}

\begin{figure*}
\centerline{\psfig{figure=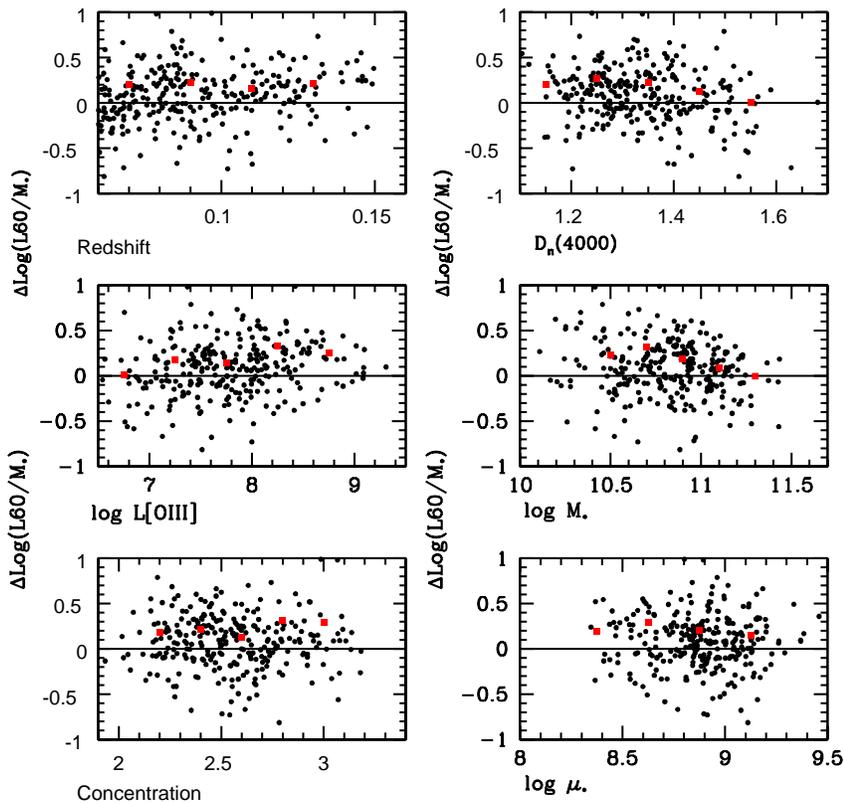,width=1.35\hssize}}
\caption{The difference between the  normalized 60 $\mu$m IR luminosity 
of an AGN of Type 2 and a matching normal star-forming galaxy is 
plotted for a sample of 254
galaxy-AGN pairs. The difference is shown as a function of
the redshift, 4000 \AA\ break strength, extinction-corrected [OIII]
luminosity, stellar mass, concentration and stellar surface density
of the AGN in each pair.  Red symbols indicate the average value
of $\Delta \log L_{60}/M_*$ evaluated in bins of each of these 6 parameters.}
\end{figure*}

Figure 11 shows histograms of  the distribution of
$\Delta \log L_{60}/M_*$ , $\Delta \log L_{100}/M_*$ and $\Delta \log L_{\rm FIR}/M_*$.
Note that the results for $L_{100}$ and $L_{\rm FIR}$ are based on a sample of
111 pairs from the catalogue with  reliable flux measurements in both the 60 and
100 $\mu$m bands. Fig. 11 demonstrates the excess IR emission in AGN is larger
at 60 $\mu$m than at 100$\mu$m.

\begin{figure*}
\centerline{\psfig{figure=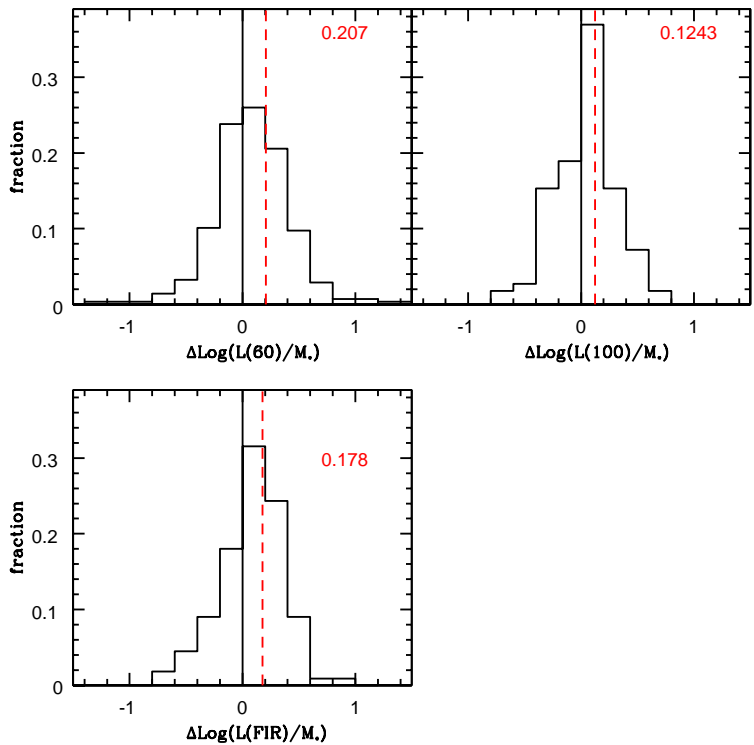,width=1.2\hssize}}
\caption{Histograms of the difference between the  normalized  IR luminosity 
of an AGN of Type 2 and a matching normal star-forming galaxy are 
plotted for  samples of 
galaxy-AGN pairs.  Results are shown for the 60 $\mu$m, 100 $\mu$m and FIR
luminosities. The average value of $\Delta \log L/M_*$  is indicated by the red dashed line
and its value is given in the top right corner of each panel. }
\end{figure*}

\section {Interpretation and Discussion}

We have demonstrated that when IRAS-selected AGN and normal
star-forming galaxies are carefully
matched in terms of parameters such as stellar mass, size, concentration, redshift and 4000 \AA\
break, the AGN exhibit ``excess'' far-IR emission af about 0.18 dex in $\log L_{\rm FIR}/M_*$.

What causes this excess  emission in AGN?
One possibility is that the excess  originates from the
active nucleus itself. The question one might then ask is whether
this is energetically feasible. In Type 2 AGN, the central engine   
is obscured and the AGN luminosity can only be estimated indirectly.
Heckman et al. (2004) have discussed how the [OIII] luminosity can be used as a
tracer of AGN activity. They used the bolometric correction to the [OIII] luminosity derived
for Type 1 AGN to estimate the average mass accretion rate onto black holes
in the local Universe.  The implied bolometric correction was $L_{Bol}/L_{O3} \sim 3500$,
where $L_{O3}$ was the [OIII] luminosity in solar units, uncorrected for extinction
due to dust. The quoted uncertainty on this conversion was $\sim 0.4$ dex.

In the left panel of Figure 12, we plot $L_{\rm FIR}$
as a function of the raw, uncorrected  [OIII] luminosity for the AGN
in our sample with  reliable flux densities at 60 and 100 $\mu$m. We have 
found that these AGN have on average a  $0.18$ dex excess in
$\log L_{\rm FIR}/M_*$. This would imply that one third of the total
FIR luminosity is from the AGN. In the left panel of Figure 12,
we see  that the median ratio of FIR to [OIII] luminosity is 5 dex. 
In the assumption that the quoted $0.18$ dex excess in $\log L_{\rm FIR}/M_*$
is entirely due to the AGN, a FIR-[OIII] luminosity ratio of 5 dex would
imply a typical ratio of FIR (AGN) to [OIII] of several 10$^4$. This
is about an order-of-magnitude greater
than the entire bolometric luminosity of the AGN if we assume the
bolometric correction of Heckman et al. At first glance, this would appear to
rule out the hypothesis  that the observed excess could be produced by the 
active nucleus. 
We caution, however, that our IRAS-selected
AGN are not typical of the general population of AGN in
the SDSS spectroscopic sample. In particular, they have considerably higher
H$\alpha$/H$\beta$ ratios, which implies that the amount of extinction
for the [OIII] line is larger. In the right hand panel of
Figure 12 we plot $L_{\rm FIR}$ as a function of the {\em extinction-corrected}
[OIII] luminosity. The average extinction correction to [OIII] 
for the AGN in our sample is a factor of $\sim 100$. 
This is five times larger than the correction derived
for the general population of  SDSS AGN with raw [OIII] luminosities
in the same range as those in our SDSS/IRAS sample. 
It is not unreasonable to suppose that strong systematic effects may
arise when applying a calibration derived for ``typical'' Type 1 AGN
to a sample of very dusty Type 2 AGN.
Figure 12 shows that if the extinction correction
is applied, the median ratio of FIR to [OIII] luminosity is $\sim$3 dex. 
The FIR excess attributed to the AGN would now be only 
several hundred times the extinction-corrected [OIII] luminosity.
This is still rather large to account for the measured infrared excess 
of AGN.

\begin{figure*}
\centerline{\psfig{figure=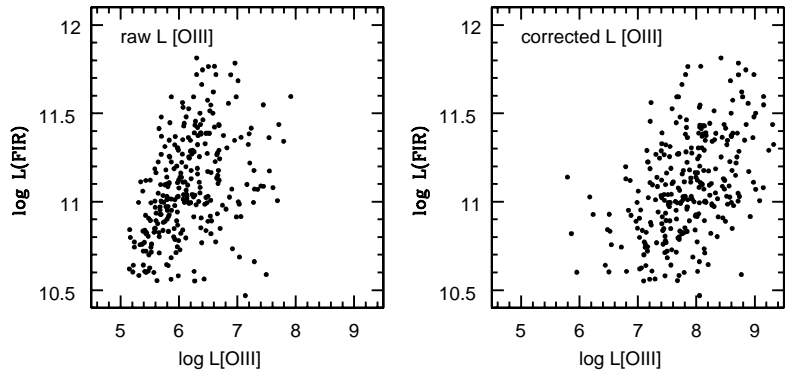,width=1.25\hssize}}
\caption{Left: The FIR luminosities of IRAS-selected AGN of Type 2
are  plotted as a function of
their raw (uncorrected) [OIII] luminosities. Right: The FIR luminosities are plotted
as a function of extinction corrected L[OIII].}
\end{figure*}

An alternative explanation is that the excess infrared emission 
is caused by  an extra component of
star formation, which is not reflected in the measured  4000 \AA\
break strengths. Kauffmann et al. (2003) showed that the distribution
of AGN with strong [OIII] emission  in the plane of H$\delta$ absorption line strength versus
4000 \AA\ break strength was systematically different than that of ordinary
star-forming galaxies of the same stellar mass. The AGN were
offset to higher values of H$\delta_A$ at a given value of D$_n$(4000),
indicating that they were more likely to have experienced a recent burst
of star formation. In the left panel of Figure 13, we compare
the H$\delta$ equivalent widths of  IRAS-selected AGN and normal
star-forming galaxies
and find that there is no significant offset in mean equivalent
width  between the two populations. Both populations are characterized by
strong H$\delta$ absorption lines and irregular morphologies
(Pasquali et al., in preparation)  and we conclude that
IRAS selection favours galaxies with bursty 
star formation histories irrespective of whether or not
they are classified as an AGN.
If there is an excess component of star formation in the AGN in
our sample, it clearly cannot be
detected using standard indicators in the optical part of the spectrum.
In the left panel of Fig. 13, we compare the Balmer decrements of
the AGN  and the normal star-forming galaxies. The AGN exhibit a 
small offset towards
larger values of $H\alpha/H\beta$. This may indicate that AGN contain
slightly more dust, but an offset in this direction is also expected
because of the different ionization conditions in these systems. 
The offset in Balmer decrement between AGN and 
normal star-forming galaxies is not
sufficient to explain the offset in FIR luminosity; we have verified
this by creating AGN/galaxy samples that are closely matched in
redshift ($< 0.03$), stellar mass ($< 0.25$ dex), D4000 ($< 0.07$) and 
H$\alpha$/H$\beta$ ($< 0.005$ dex) and we find
that the mean offset in $L_{\rm FIR}/M_*$  changes very little.
  
\begin{figure*}
\centerline{\psfig{figure=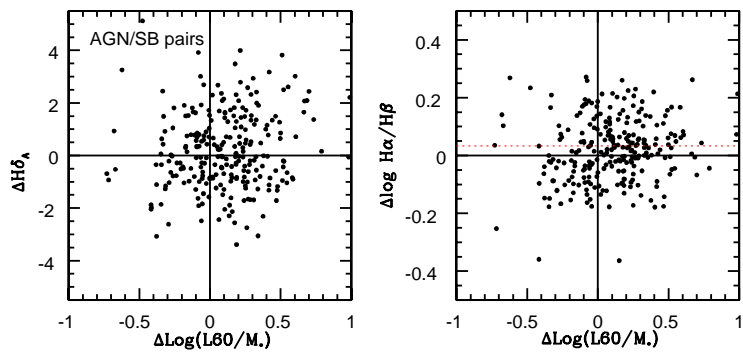,width=1.25\hssize}}
\caption
{Left: The difference between H$\delta_A$ index strength is plotted as function
of $\Delta L_{60}/M_*$ for matched galaxy/AGN pairs.
Right: The difference in the logarithm of the Balmer decrement is plotted
as a function of $\Delta L_{60}/M_*$ (the red dotted line indicates the
average value of the difference).}  
\end{figure*}

\begin{figure*}
\centerline{\psfig{figure=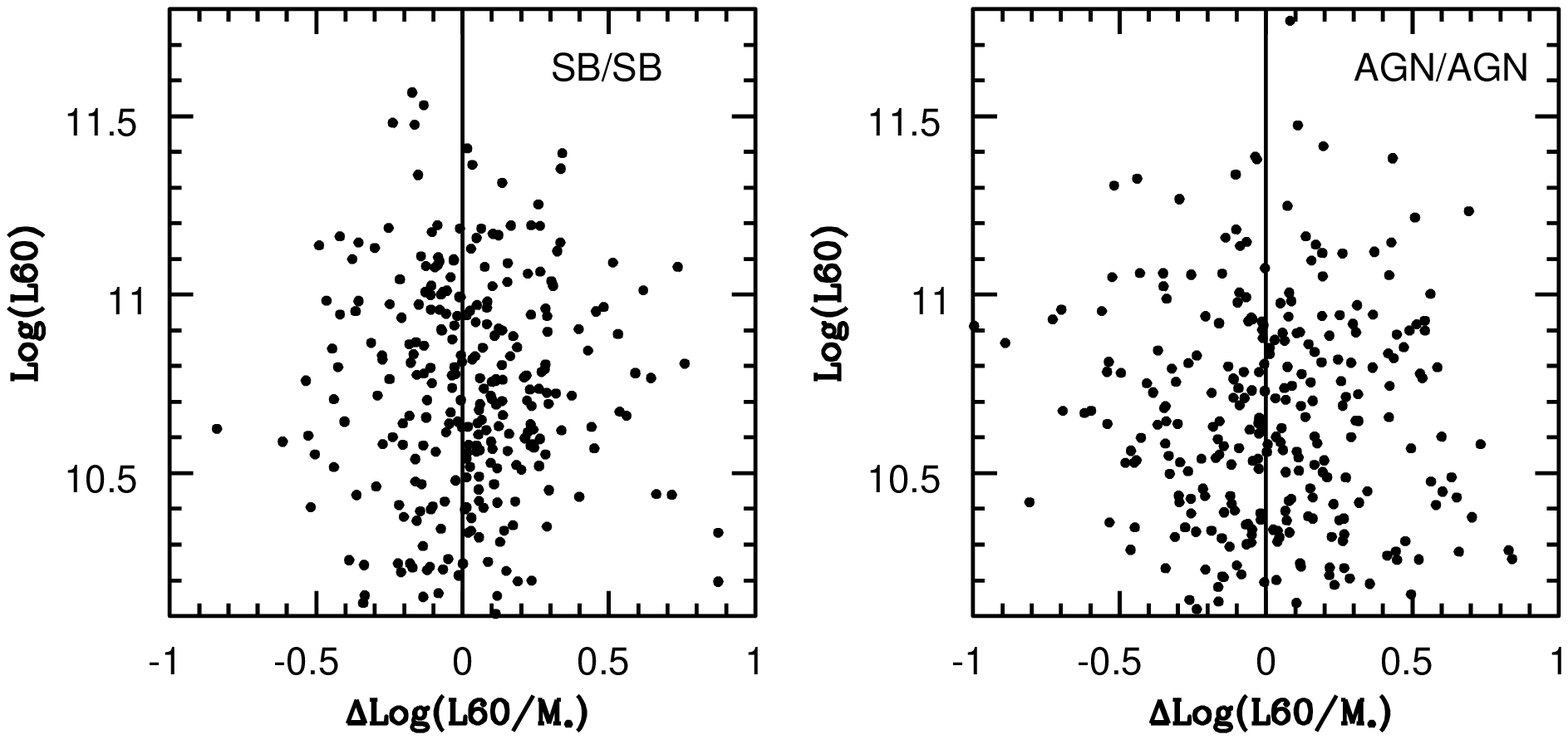,width=1.25\hssize}}
\caption{The difference in normalixed 60 $\mu$m luminosity is plotted as a function of
L(60) for a sample of matched pairs of normal star-forming galaxies (left) and 
AGN of Type 2 (right).} 
\end{figure*}

At infrared wavelengths, the only diagnostics available to us are the IRAS
60 and 100 $\mu$m fluxes. It is possible to study whether AGN exhibit a greater
degree of {\em scatter} in their far-IR properties relative to 
normal star-forming galaxies. We have created
samples of AGN/AGN and galaxy/galaxy pairs using the exact same matching
criterion used for the galaxy/AGN pairs. We then compare                    
the scatter in 60 $\mu$m luminosity differences for the two pair samples.   
Figure 14 presents the results of this analysis
and shows clearly that the AGN exhibit greater spread in their IR
luminosities than the star-forming  galaxies in our sample.
This is consistent with the idea that the star formation in AGN is
more episodic but it does not prove the hypothesis that an excess
population of young stars is present in the AGN. One could imagine that
dust heating  by the central source might  also be subject to temporal fluctuations. 
What is  required in order to understand our results in more detail  is  
high resolution imaging of these galaxies at infrared or radio wavelengths.
This would enable us to 
compare the distributions of the dust emission and star-forming sites
in the two classes of galaxies.
With the combination of spatial information and good statistics, it should
be possible to disentangle the physical mechanisms responsible
for the excess far-infrared emission observed in nearby Seyfert 2 galaxies. 
\vspace{0.4cm}  

\section*{Acknowledgments}
We would like to thank S. Charlot and C. Tremonti for helpful discussions,
and an anonymous referee for  useful comments that improved the paper.
\par\noindent
Funding for the creation and distribution of the SDSS Archive has been provided
by the Alfred P. Sloan Foundation, the Participating Institutions, the National
Aeronautics and Space Administration, the National Science Foundation, the U.S.
Department of Energy, the Japanese Monbukagakusho, and the Max Planck Society.
The SDSS Web site is http://www.sdss.org/.
\par\noindent
The SDSS is managed by the Astrophysical Research Consortium (ARC) for the
Participating Institutions. The Participating Institutions are The University
of Chicago, Fermilab, the Institute for Advanced Study, the Japan Participation
Group, The Johns Hopkins University, Los Alamos National Laboratory, the
Max-Planck-Institute for Astronomy (MPIA), the Max-Planck-Institute for
Astrophysics (MPA), New Mexico State University, University of Pittsburgh,
Princeton University, the United States Naval Observatory, and the University
of Washington.

\end{document}